# Parametric Sensitivity Analysis of a Thermal Test Cell Model Using Spectral Analysis

Thierry Alex MARA
University of La Réunion Island, Laboratoire de Génie Industriel, BP 7151, 15 avenue René Cassin, 97 705 Saint-Denis, France. email : mara@univ-reunion.fr

Harry BOYER
University of La Réunion Island, Laboratoire de Génie Industriel, BP 7151, 15 avenue René Cassin, 97 705 Saint-Denis, France. email : harry.boyer@univ-reunion.fr

François GARDE
University of La Réunion Island, Laboratoire de Génie Industriel, BP 7151, 15 avenue René Cassin, 97 705 Saint-Denis, France. email : garde@univ-reunion.fr

**Abstract** : This paper deals with a new method to perform parametric sensitivity analysis. Such a study is very important for modellers because it can provide fruitful information. Indeed, it can point out model's weaknesses and allows to identify the most important parameters in the model, which modellers must know accurately to provide reliable predicted results. After describing the approach, an application of the method in building thermal simulation is discussed. The study concerns a real cell-test and gives coherent results as some of the most influential factors can be physically interpreted. Moreover, results from this analysis allow us to plan the design of next experiments.

## Introduction

The improvement of building thermal behaviour is a very important challenge because of the electrical consumption. The use of building thermal simulation software is necessary to achieve this task. But, before using such a program, one must ensure that its results are reliable. To do so, a methodology must be applied including the verification of numerical implementation and experimental validation. At University of La Réunion Island, we developed a building thermal simulation software and we would like to compare its results to measurements. But, before carrying out experiments on a test-cell, we would like to diagnose it to better plan experiments through the use of sensitivity analysis methods.

Sensitivity analysis (or SA) of model output is a very important stage in model building and analysis. It's applied in simulation studies in all kinds of disciplines : chemistry (Robin , 1998), physics (Adebiyi, 1998), management science (Balson, 1992), and so on. In building thermal simulation field, SA is more and more applied (Lomas & Eppel 1992**,** Fürbringer 1994**,** Rahni 1998, Aude 1998). Indeed, SA can help increase reliability in building thermal simulation software's predictions. The purpose of this paper is to introduce an easy method to identify the most influential factors and evaluate their effect. An application is given in thermal building showing results that can be physically interpreted (reinforcing the reliability of the method) and pointing out the weaknesses of the model. These results are helpful as it gives information on how to plan future experiments.

**Nomenclature**

| | | | |
|---|---|---|---|
| **Variables** | | **Thermal properties** | |
| $T_{sky}$ (°K) | Fictive sky temperature | $e$ (m) | Thickness |
| $T_{env}$ (°K) | Fictive environment temperature | $\lambda$ (W/m².K) | Conductivity |
| $T_{ao}$ (°K) | Outdoor air temperature | $\rho$ (Kg/m$^3$) | Density |
| $\varphi_{lwo}$ (W/m²) | Outdoor short-wave heat flux radiation density | $c$ or $C_p$ (J/Kg.K) | Specific heat at constant pressure |
| $T_{so}$ (°K) | Outdoor surface temperature | $\tau$ | Transmittance |
| | | $\alpha_i$ & $\alpha_o$ | Indoor and outdoor surface absorptance |
| **Signal's characteristics** | | $K$ (W/m².K) | Thermal conductance |
| $a_h$ | Fourier's coefficients at frequency $f_h$ | Ce & Ci (J/m².K) | Outside node and inside node thermal capacities |
| $\Gamma_X(f)$ | Power Spectral Density of signal x | *Hci* & *Hco* (W/m².K) | Indoor and outdoor convective heat transfer coefficients |
| $\delta$ | Dirac Function | *Hrc* (W/m².K) | Outdoor surface radiative heat transfer coefficient with fictive sky temperature including view factor |
| $\sigma^2_x$ or var(x) | Variance of variable x | *Hre* (W/m².K) | Outdoor surface radiative heat transfer coefficient with environment including view factor |

# I. The proposed methodology

Parametric Sensitivity Analysis can be regarded as a study of error propagation in models. So, a naturally way to perform such a work is simply propagate information in the model via model's factors and verify if this information is present in the outputs. In which case, one would infer that parameters associated to the information found are influential. Moreover, it could be possible to evaluate the influence of each of them according to the intensity of their information in the output. This last point is very important because that allows to determine *sensitivity indices* that would allow to quantify the influence of each parameter and to identify the most important one.

The problem that crops up is then the following : what information could be associated with the different parameters? An idea is to make each parameter vary as a *sinusoid* (i.e. in a periodic manner) so the associated information is a frequency. Thus, this information can easily be found by calculating Fourier Transformed (FT) or Power Spectral Density (PSD) of the predicted results (cf. figure 1). In fact, this approach is a particular case of a more general SA technique called FAST (Fourier Amplitude Sensitivity Test) developed by CUKIER & al [1973].

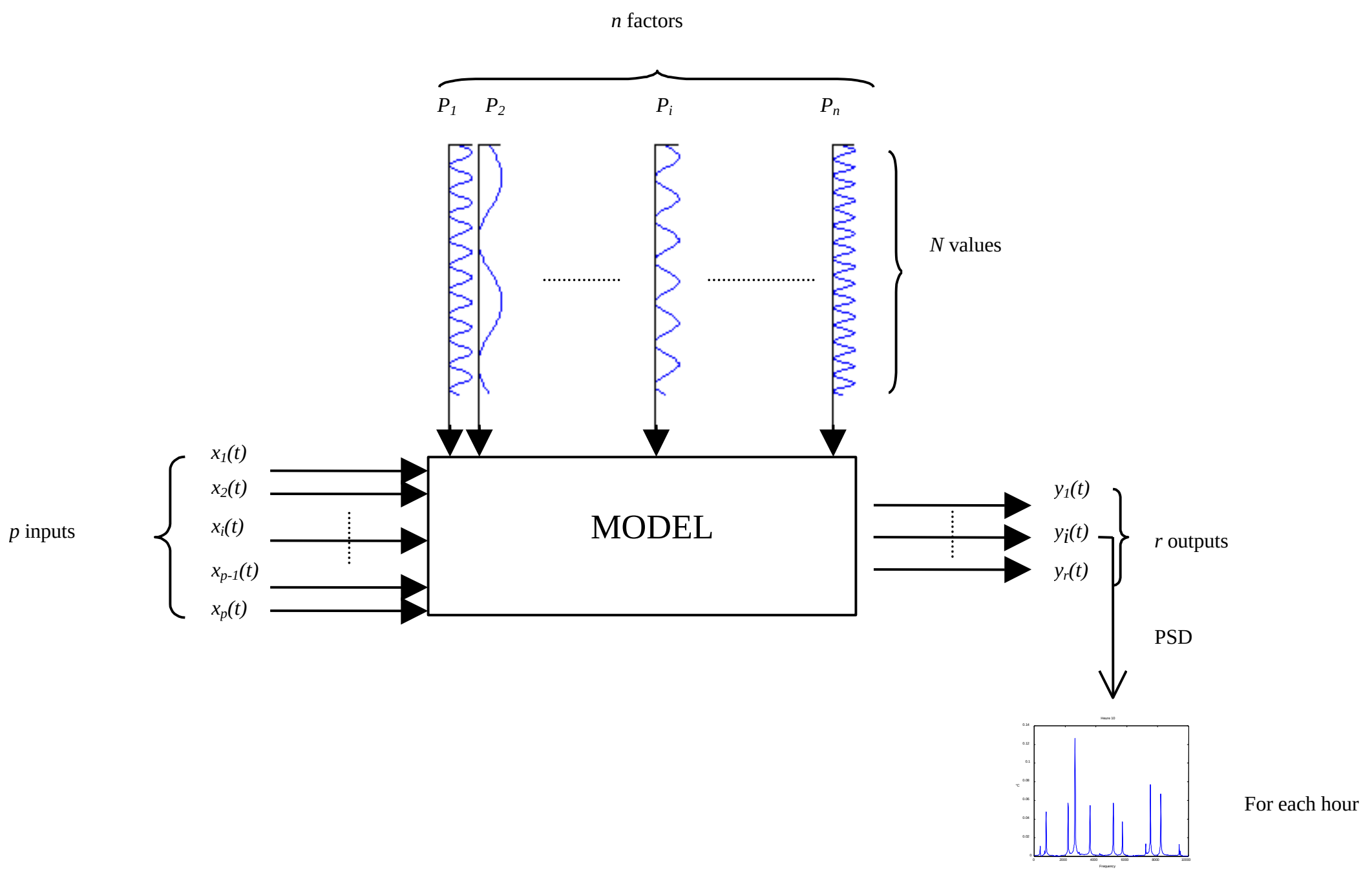

*Figure 1*

## *How do we use the method in practice and what results are expected?*

Let's consider a k parameter model $Y = F(X_1, X_2, .., X_h, .., X_k)$ and let's perform simulations by varying each factor as a *sinus* with different frequency so that each factor can be written as:

$X_h = X_{base}.(1+d_{\%}.\sin(2.\Pi.f_h)) \Rightarrow X_h \in [X_{base} - d_{\%}.X_{base} \quad X_{base}+d_{\%}.X_{base}\ ]$

where $X_{base}$ denotes the base case combination factors and $0 < d_{\%} < 1$.

In case of a linear model, one would expect that frequencies assigned to factors would be found in the vector $\Delta Y = Y_{base} - Y$

where $Y$ is the predicted vector of the different simulations.

$Y_{base}$ is the predicted output obtained when $X_h = X_{base}$

which means that $\Delta Y$ would be a superposition of sinuses[1], so we'd find :

$$\Delta Y = \sum_{h=1}^{p} a_h . sin(2\pi . f_h)$$

where p≤k depends on the factors' influence.

But if the model contains second order effect (not linear) between factors, we'd find :

$$\Delta Y = \sum_{h=1}^{p} a_h . sin(2\pi . f_h) + \sum_{h=1}^{q} \sum a_{hh'} . sin(2\pi . f_h) . sin(2\pi . f_{h'})$$

$a_h$ and $a_{hh'}$ measure on the importance of factor $h$ and its interactions respectively.

The FT of $\Delta Y$ is :

$$FT_{\Delta y}(f) = \sum_{h=1}^{k} \frac{a_h}{2} . \delta(f - f_h) + \sum_{h=1}^{k} \sum_{h'=1}^{k} \frac{a_{hh'}}{2} . \left[ \frac{1}{2} \delta\left(f - \left(\left| f_h - f_{h'} \right|\right)\right) + \frac{1}{2} \delta\left(f - \left(f_h + f_{h'}\right)\right) \right] \quad \forall f_h \geq 0 \qquad (3a)$$

with

$FT_{\Delta y}(f)$ is the FT of $\Delta Y$

$a_h$ & $a_{hh'}$ are the Fourier coefficients.

If, instead of calculating the FT of $\Delta Y$, one calculates its PSD, equation (3a) becomes :

$$\Gamma_{\Delta y}(f) = \sum_{h=1}^{k} \frac{a_h^2}{2} . \delta(f - f_h) + \sum_{h=1}^{k} \sum_{h'=1}^{k} \frac{a_{hh'}^2}{2} . \left[ \frac{1}{2} \delta\left(f - \left(\left| f_h - f_{h'} \right|\right)\right) + \frac{1}{2} \delta\left(f - \left(f_h + f_{h'}\right)\right) \right] \qquad \forall f_h \geq 0 \qquad (3b)$$

where the first sum represents main effects and the second sum is the second order effects, that represents interactions between 2 parameters.

So, we can see that Fourier coefficients would give the information of the relatively importance of each factor and their interactions. Moreover, we know that $\sigma_{\Delta y}^2 = \int_{-\infty}^{+\infty} \Gamma_{\Delta y} df$ so we obtain :

[1] *In fact, $\Delta Y$ may contain Cosines but to make our presentation not cumbersome, we didn't take Cos into account as it doesn't change the information expected (frequencies).*

$$\sigma^2_{\Delta y} = \int_0^{+\infty}\left(\sum_{h=1}^{k} a_h^2.\delta(f - f_h) + \sum_{h=1}^{k}\sum_{h'=1}^{k} a_{hh'}^2.\left[\frac{1}{2}\delta\left(f - \left(\left|f_h - f_{h'}\right|\right)\right) + \frac{1}{2}\delta\left(f - \left(f_h + f_{h'}\right)\right)\right]\right)df \text{ as } \Gamma_{\Delta y} \text{ ispair}$$

$$\Leftrightarrow \sigma^2_{\Delta y} = \sum_{h=1}^{k} a_h^2 + \sum_{h=1}^{k}\sum_{h'=1}^{k} a_{hh'}^2$$

$$\Leftrightarrow 1 = \sum_{h=1}^{k}\frac{a_h^2}{\sigma^2_{\Delta y}} + \sum_{h=1}^{k}\sum_{h'=1}^{k}\frac{a_{hh'}^2}{\sigma^2_{\Delta y}} \qquad (4)$$

which is the variance decomposition of the output variation due to each factor.

*Comments :*

To use such a technique it is necessary to perform N simulations, depending on the number of factors, to find each frequency in the spectrum. Moreover, to avoid frequencies superposition one must have a set of incommensurate frequencies. Interactions between factors induce additional frequencies of the same amplitude. In equation (3b), second order effects induce 2 frequencies which are $|f_h-f_{h'}|$ et $(f_h+f_{h'})$. It can be inferred that $p^{th}$-order interactions induce $2^{p-1}$ frequencies which are $\sum_{i=1}^{p} f_i$ plus linear combinations of the form $\left|\sum_{i \neq j}^{p} f_i - f_j\right|$ $\forall$ j $\in$ [1 , p]. One can notice that the amplitude at each frequency (which is the squared Fourier coefficient) is a measure of the importance of each parameter. More precisely, the ratio $\frac{a_h^2}{\sigma^2_{\Delta y}}$ is the amount of the variance of $\Delta y$ due to factor $h$.

# II. Application to the thermal model of a real Test Cell

The survey concerns a real test cell that was erected at University of Reunion Island for experimental validation of building thermal airflow simulation software (see Garde, 1997). After describing the *building* and some model assumptions, we'll apply the methodology previously introduced to identify the most important factors in the model. We recall that such a study is helpful to guide future experiments.

## *II.1 Real building description*

The studied test cell is a cubic-shaped building with a single window on the South wall and a wooden door on the North. All vertical walls are identical and are composed of cement fibre and polyurethane, the roof is constituted of steel, polyurethane and cement fibre and the floor of concrete slabs, polystyrene and concrete. The building considered is well-isolated. Base values of each layer's thermal properties are regrouped in Table 2 (cf. Appendix). Picture 1 shows a picture of the cell-test, and on the left, the weather station that provides solicitations (inputs) to our building model.

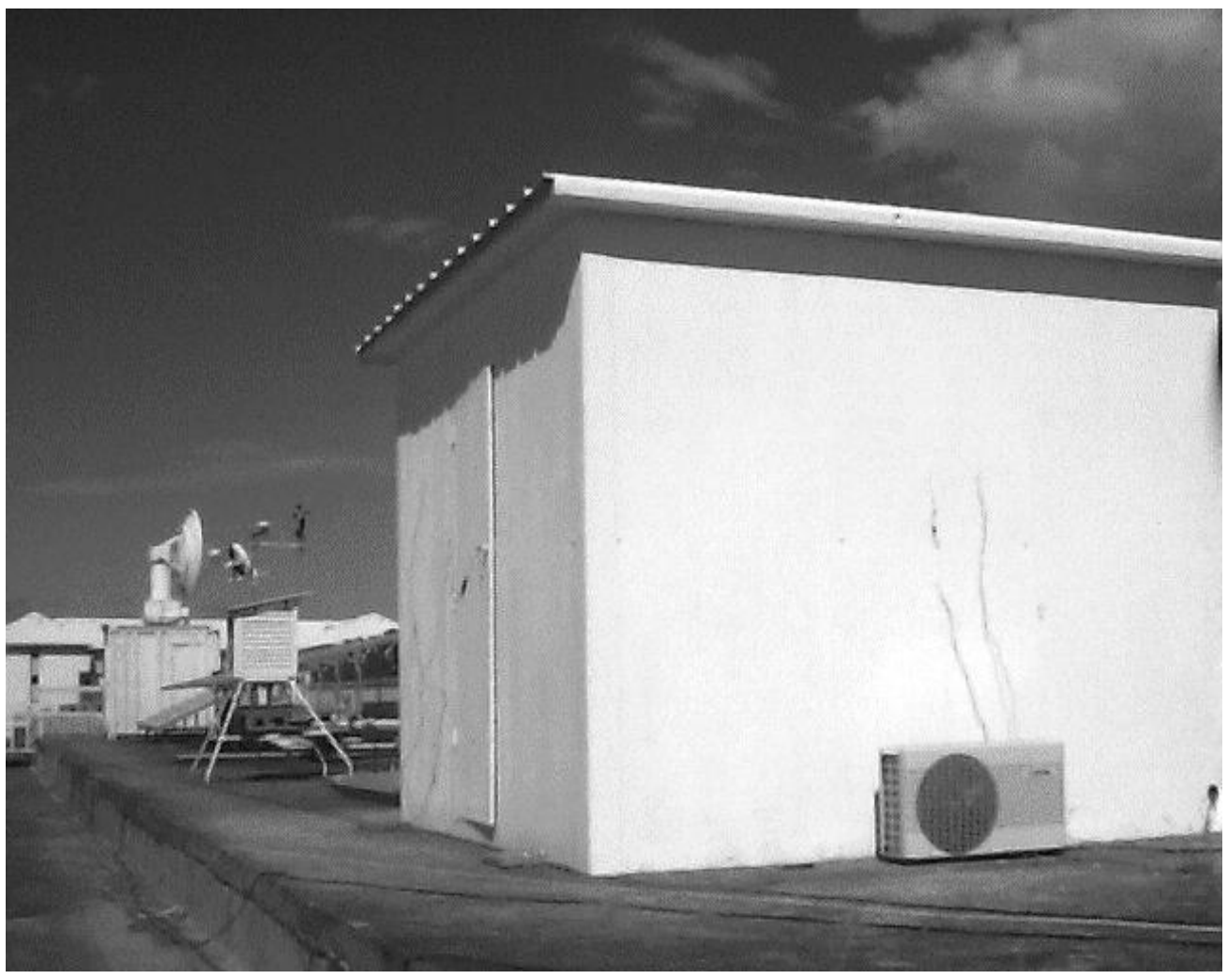

*Picture 1 : Picture of the test-cell from North-West*

The case discussed in this paper is a passive one by powering off the split system we can see on the picture.

### *II.2 Model description*

A lumped approach (Boyer, 1996) is used to represent the building. It is based on the analogy between the equation of conduction of Fourier and Ohm's law. Such a model leads to a system of equations, called state equations, which in the matrix formalism has the following form :

$$C.\dot{T} = A.T + B$$

where :

*A* is the state matrix;

*B* is the solicitations matrix;

*C* is the capacities matrix;

*T* is the state vector (temperature) composed of temperatures of lumped elements;

$\dot{T}$ is the derivative of *T*.

In this survey, we consider the electrical/thermal analogy representation of heat transfer conduction through walls (cf. Scheme 1) which consists in discretizing a wall with 3 nodes by layers.

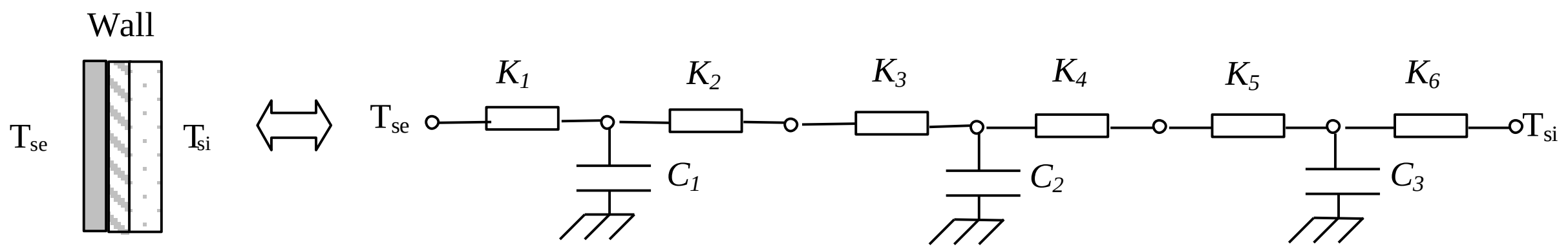

*Scheme 1 : Wall spatial discretization and representation*

The thermal capacities ($C_1,C_2,C_3$) and the thermal conductance ($K_1,K_2,K_3,K_4,K_5,K_6$) of a layer are respectively function of its thickness, specific heat, density ($e,c,\rho$) and thickness, conductivity ($e,\lambda$).

<u>*Assumptions*</u> :

Nodal analysis assumes that heat conduction transfer through walls are mono-dimensional. Indoor radiant heat transfer is linearized and the radiative exchange coefficients are identical for each wall. For outdoor long-wave radiative heat exchange, we use the following model :

$$\varphi_{lwo} = Hrc.(T_{sky} - T_{so}) + Hre.(T_{env} - T_{so}) \qquad \text{with } T_{sky} = T_{ao} - 6$$

$$T_{env} = T_{ao}$$

Fictive sky temperature ($T_{sky}$) is rarely measured. A correlation usually used, for tropical climate, to model it is the proposed one (see [Garde, 1997] or [Baronnet, 1985]). In the same way, environmental temperature is usually considered as equal to ambient temperature.

Indoor and outdoor convective exchange coefficients are also constant for each wall. (see Table 3 and 4 in Appendix for base values)

Heat flux under the floor is null. This latter assumption is reasonable here as the floor is thermally decoupled from the ground.

## *II.3 Parametric sensitivity analysis*

We distinguished all the factors even if they are identical except for *Hrc*, *Hre* and *Hri*. Thus, for example, thermal properties of cement fibre are distinguished from one wall to another and the cement fibre of the interior layer varied differently than the one of the exterior layer. In the same way, inner convective exchange coefficient (*Hci*) is distinguished from one wall to another and so on. The drawback is the increase of factors but it allows to find which factors of which wall are influential. We didn't take into account air properties which are assumed to be known accurately. This way of proceeding generated 120 factors which require 120 different frequencies.

We performed 1024 simulations by making each factor vary as a *sinusoid* ranging ±10% with respect to its base value (cf. § 1). Weather data concerns hot season when direct solar radiation passes through the south window.

In the following study, we are looking for the most important parameter for the predicted indoor air temperature. So, once the simulations are performed, we calculate the power spectrum density of

$\Delta T_i = T_{i,base} - T_{i,evol}$

where $T_{i,base}$ is the indoor air temperature obtained with the factors base value at time *i*

and $T_{i,evol}$ the indoor air temperature obtained with the different simulations at time *i*.

## Results

Figures 1 to 8 represent the spectra of the PSD of $\Delta T$ ($\Gamma_{\Delta T}(f)$) at different hours. The spectra show that there are only a few important frequencies which means that only a few parameters are influential. The analysis of the spectra (figure 1 to 10) and Table 1 show that the most influential factors are the windows properties that's to say its area (frequency 1826) and its transmittance (frequency 8435) and *Hrc* (frequency 2058). Moreover, we note that the effect of some parameters depends on the hour of the day. For instance we can notice that frequency 5433 (area of the floor) is high during day time and progressively disappears in the night. The level of the peak of a frequency at a given time gives a quantitative information about the influence of the parameter. For instance, at 1 h (Fig. 1) a variation of 10% of the outdoor radiative heat transfer coefficient (*Hrc*) will make the indoor air temperature vary from ± $0.01^{1/2}$ = ± 0.1°C. So, to evaluate the effect of each parameter, one can look at the level of its assigned frequency or one can use equation (4).

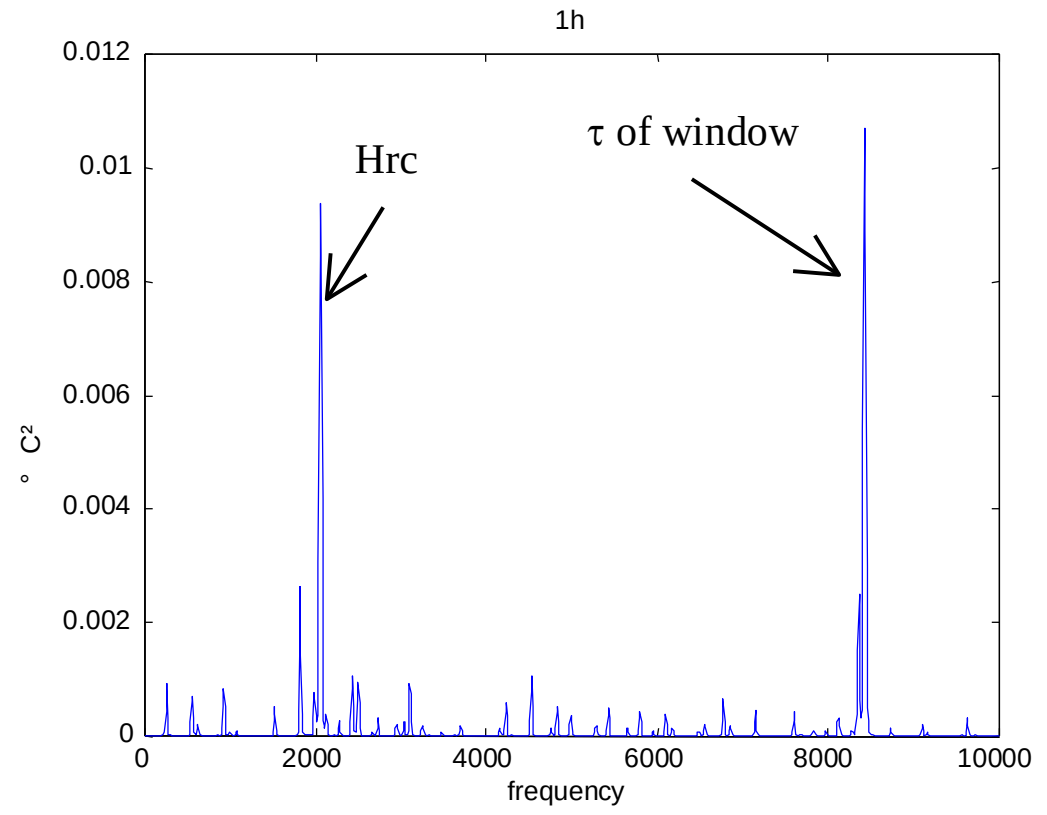

*Figure 1*

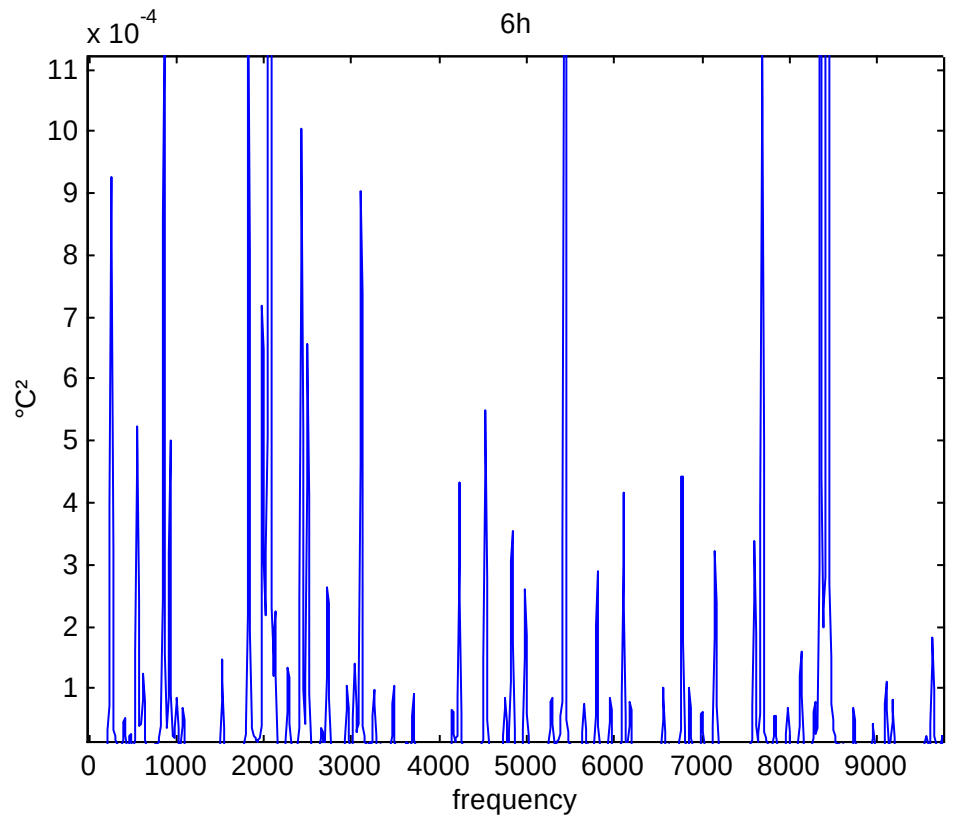

*Figure 2 : Spectrum zoomed*

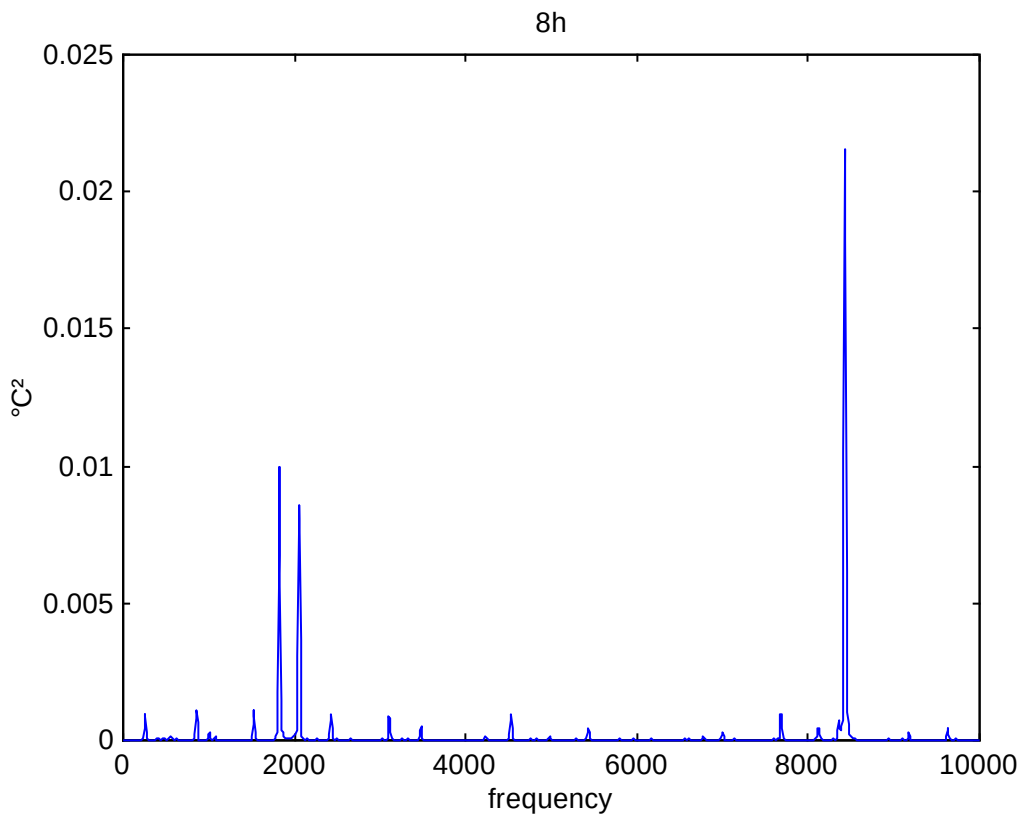

*Figure 3*

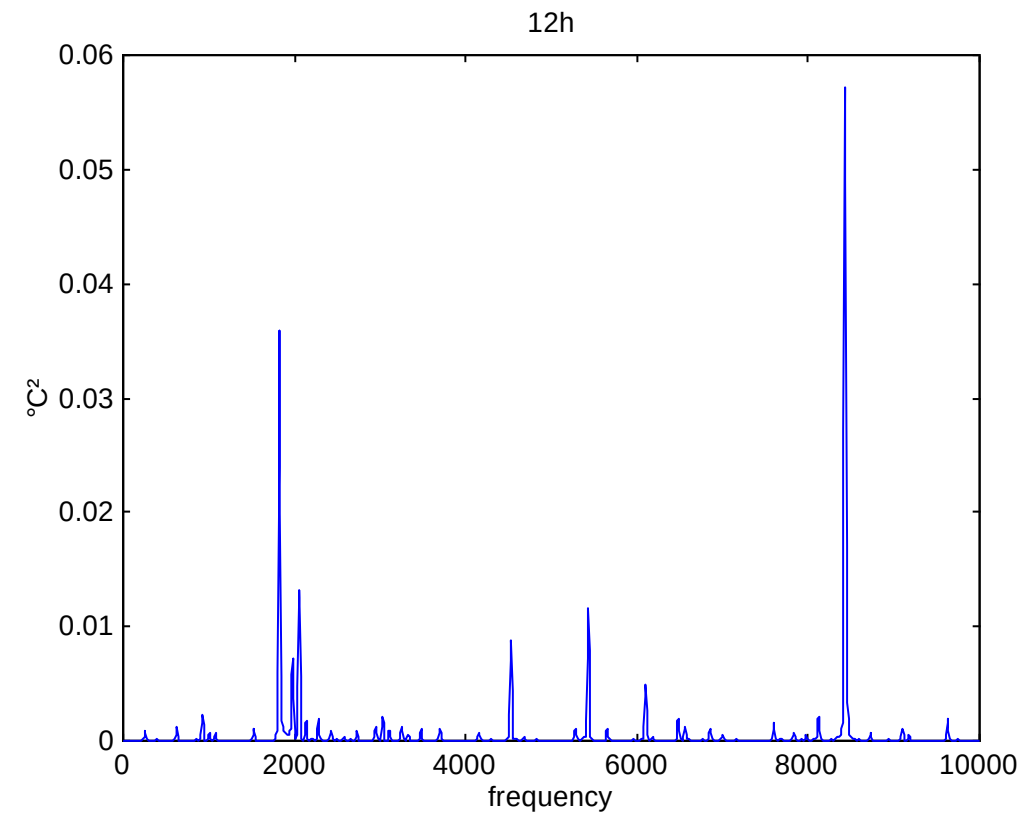

*Figure 4*

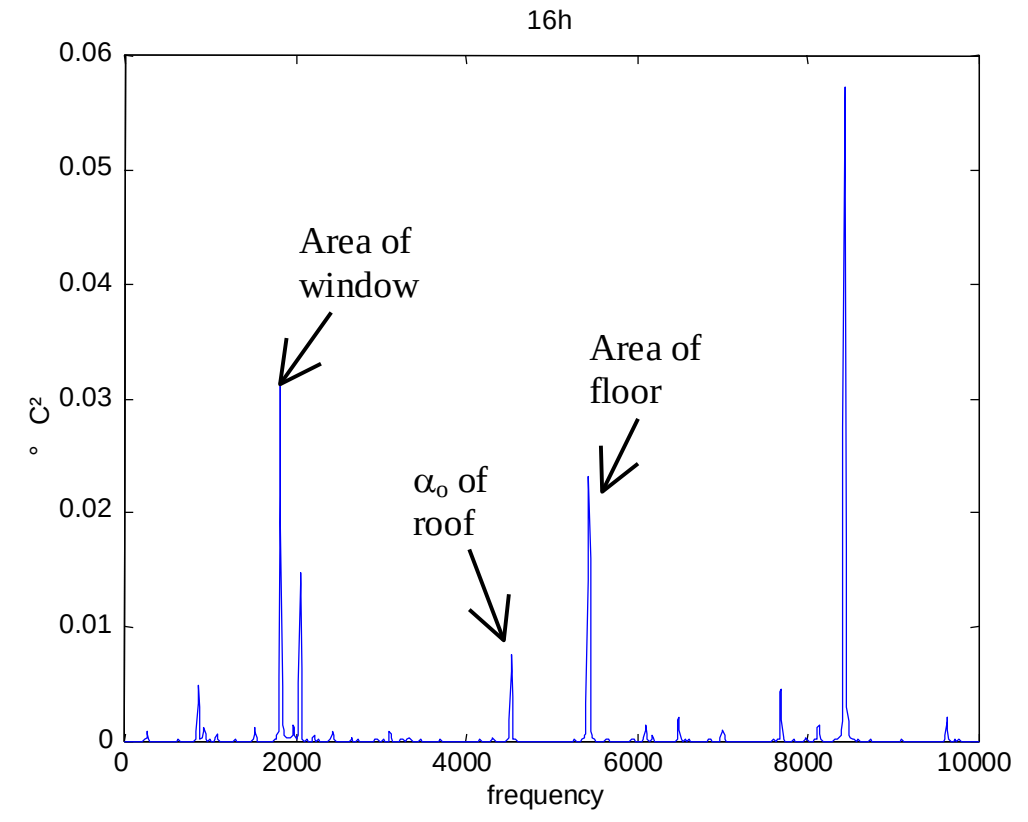

*Figure 5*

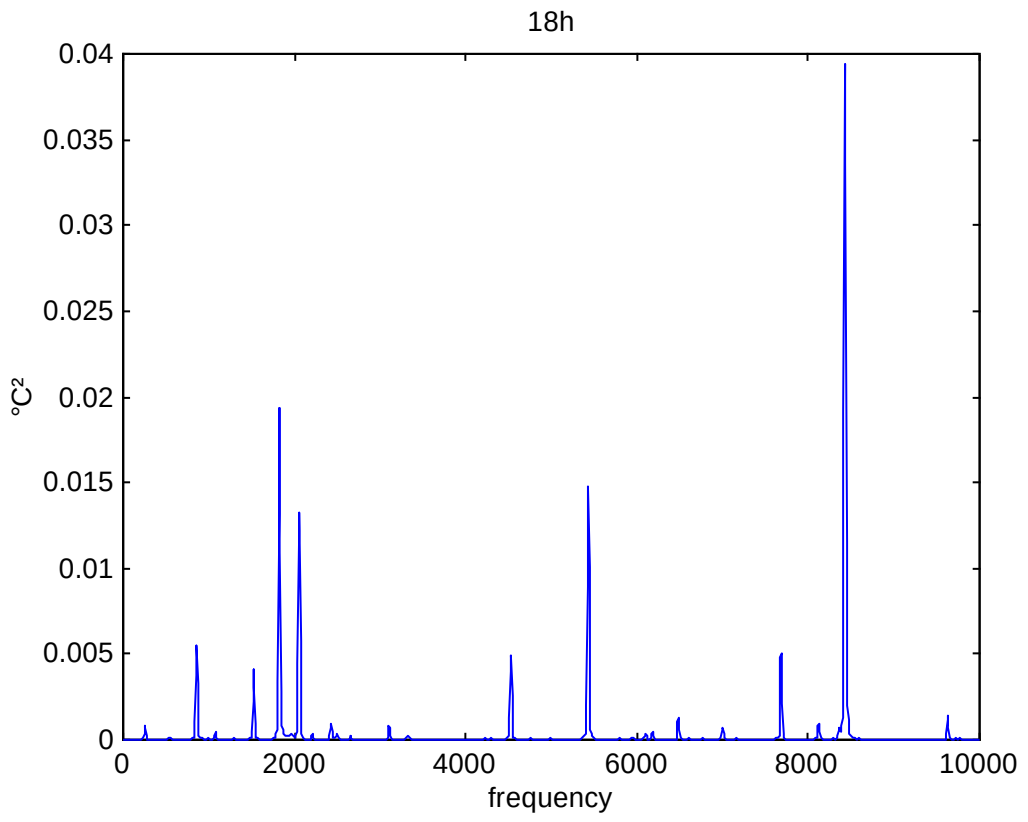

*Figure 6*

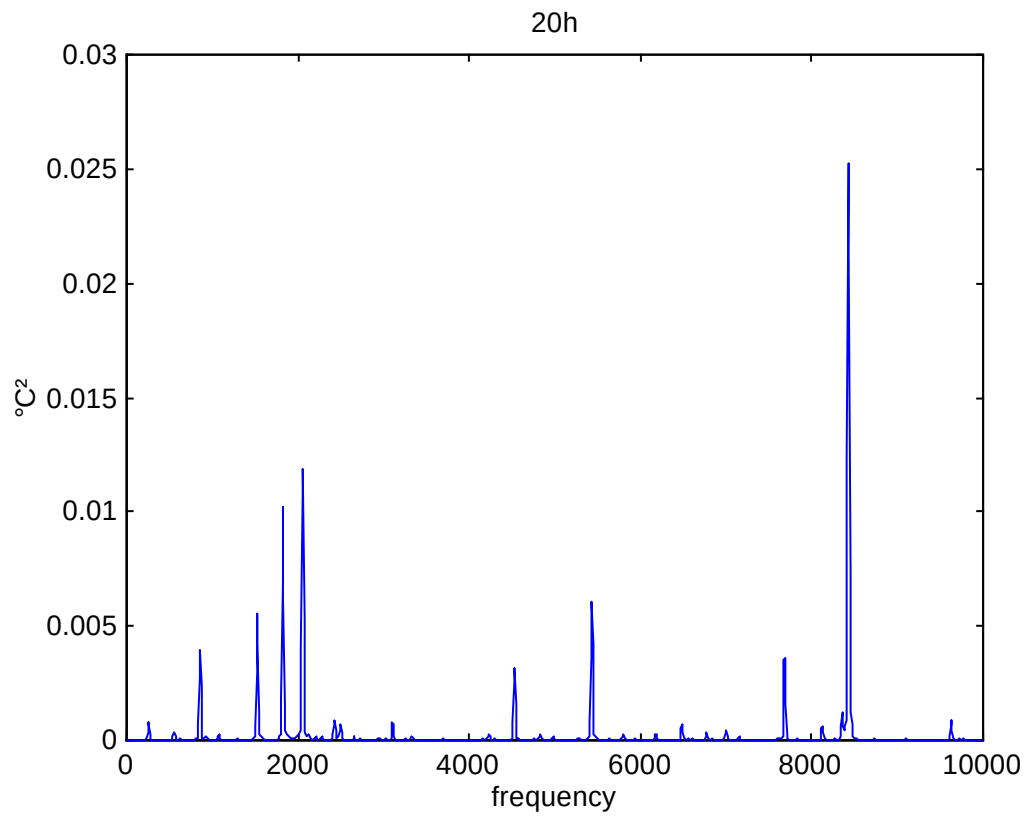

*Figure 7*

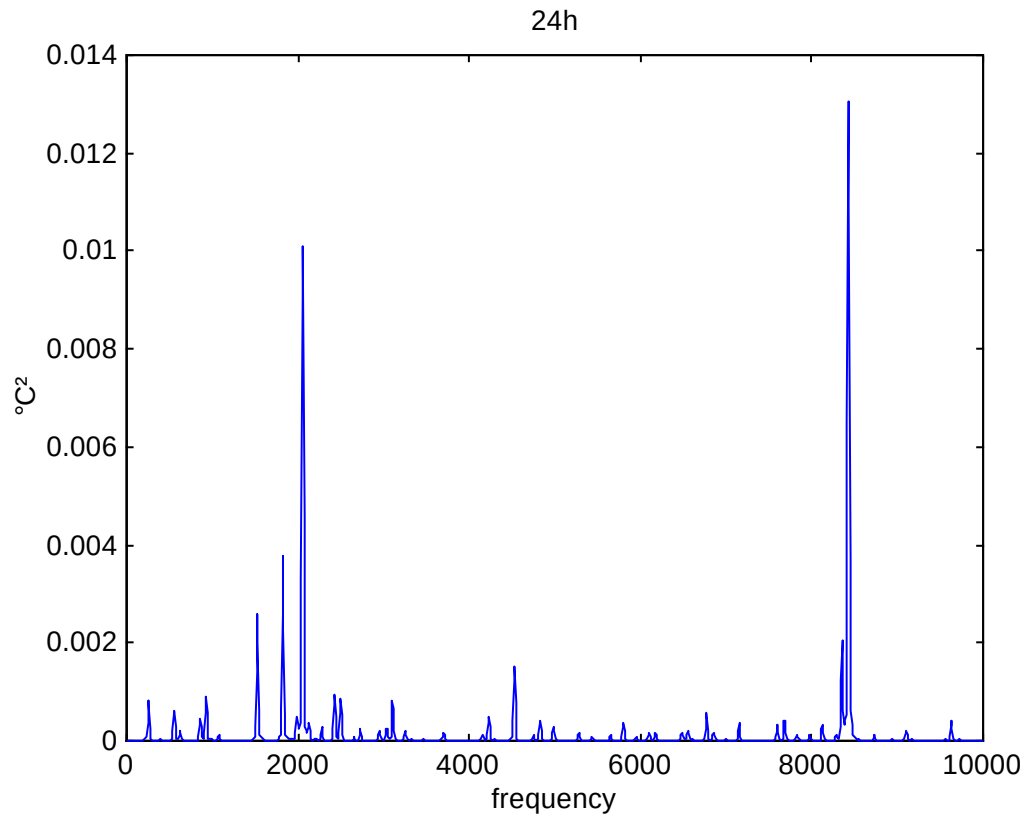

*Figure 8*

:

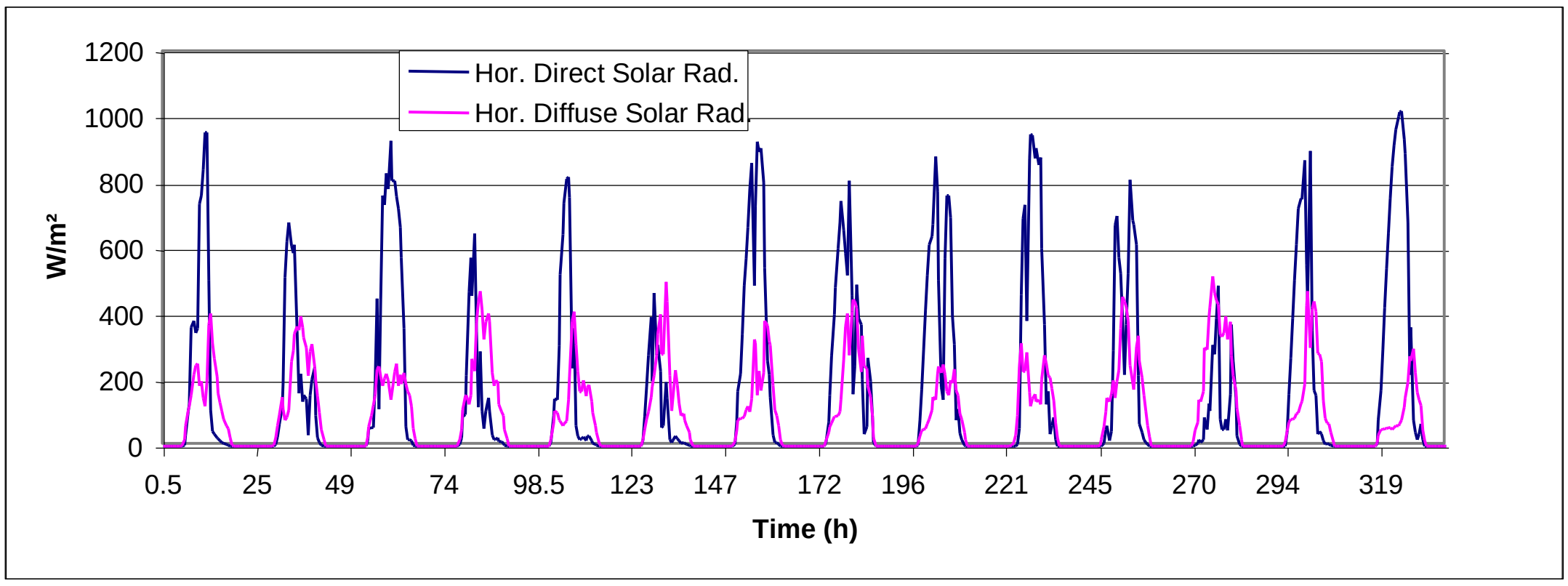

*Figure 9 : Evolution of Horizontal Direct and Diffuse Solar Radiation during the 14 days of simulations.*

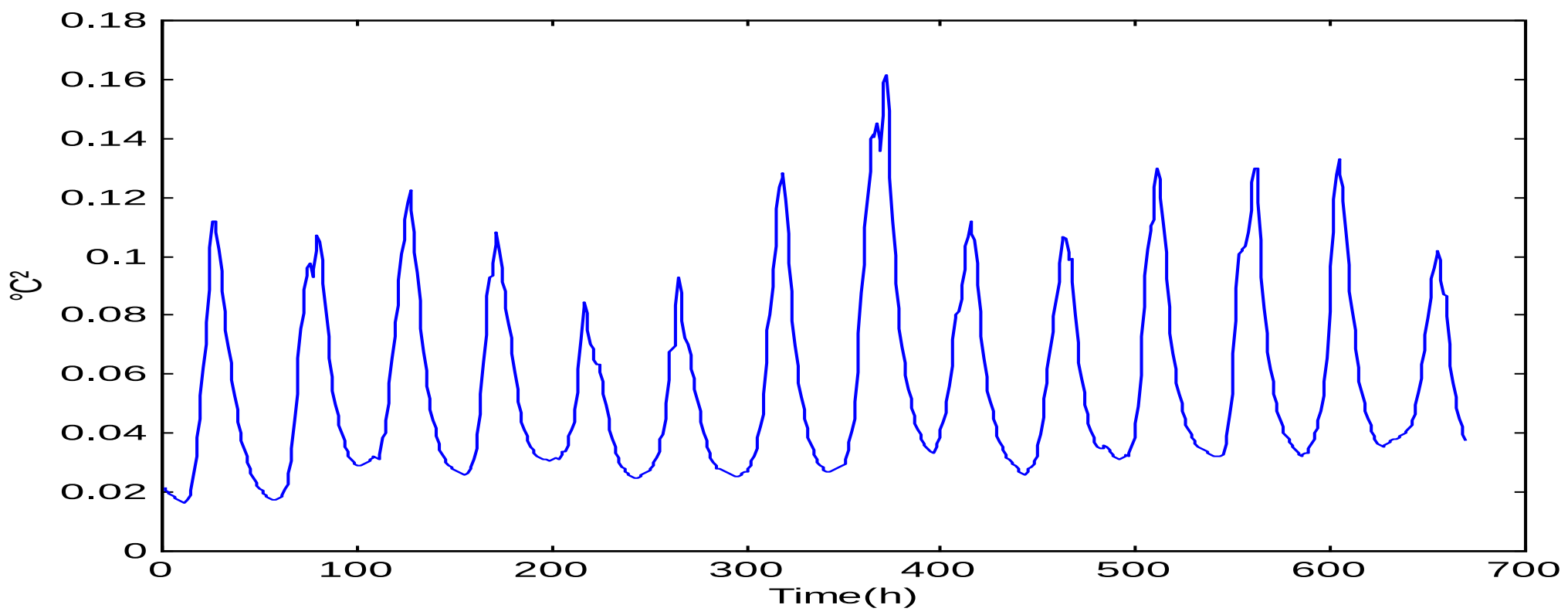

*Figure 10: Hourly evolution of var(ΔT)*

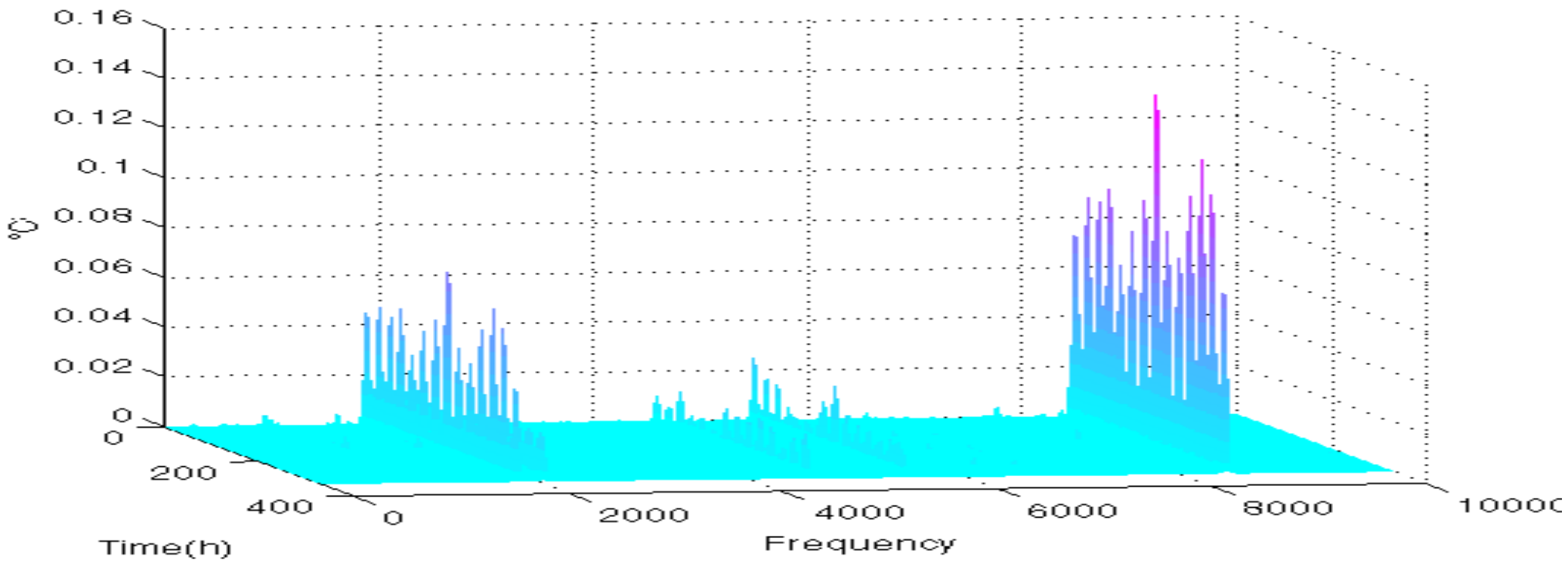

*Figure 11 :Evolution of each parameter's influence on 24 h*

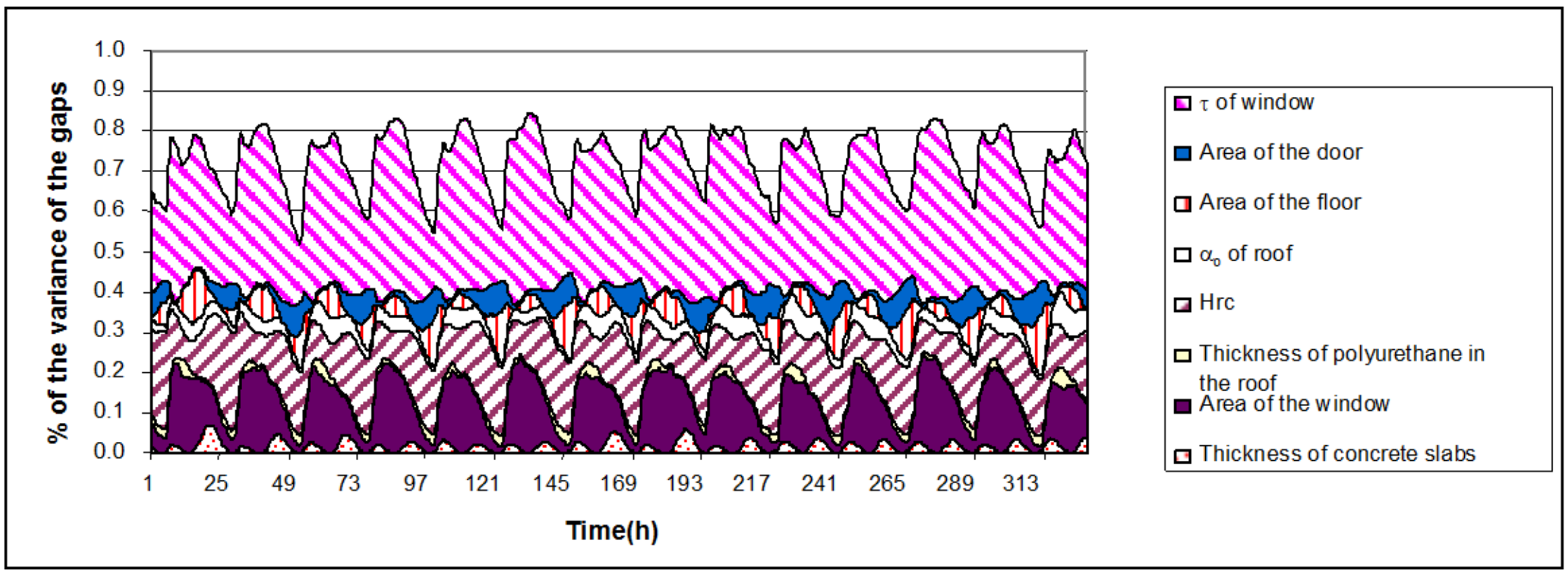

*Figure 12 : Effect of the 8 most important factors*

From eq 4, we determined the influential factors by taking into account only those who explained more than 1% of var($\Delta T$) at a given time. We found 34 parameters (see Table 1).

| Wall | Material | $e$ | $\lambda$ | $C_p$ | $\rho$ | $\alpha_e$ | *Area* | *Hrc* | $Hc_o$ | $\tau$ |
|---|---|---|---|---|---|---|---|---|---|---|
| East | Polyurethane | 7601 | 2730 | | | 8127 | | 2058 | | |
| South | Polyurethane | 4834 | 7153 | | | | | 2058 | | |
| West | Polyurethane | 6776 | 4230 | | | 6180 | 3476 | 2058 | | |
| North | Polyurethane | 5802 | 4983 | | | | | 2058 | | |
| Roof | Cement fibre | 2279 | | 3030 | 2134 | 4531 | 7002 | 2058 | 6484 | |
| | Polyurethane | 1984 | | 6105 | | | | | | |
| Floor | Concrete slabs | 1526 | 928 | 859 | 7685 | | 5433 | | | |
| | Weight concrete | 258 | | 2430 | 3110 | | | | | |
| Door | Wood | 2502 | 552 | | | 9631 | 8353 | 2058 | | |
| Window | Glass | | | | | | 1826 | 2058 | | 8435 |

*Table 1 : The influential factors and their associated frequency.*

All the spectra can be regrouped in one graph that represents the evolution of the spectra versus time (Figure 11 ). This latter shows the preponderance of frequencies 855 and 2058 and that their effect on indoor air temperature is different from one day to another.

Figure 12 shows, hour by hour, the amount of the variance of $\Delta T$ due to the most influential parameters those who explained more than 5% of *var*($\Delta T$) at a given time. These 8 parameters explain between 60 to 80% of the total variance of the gaps. The remaining amount should be explained by the low effects of the 112 other factors and interactions. The amount of the variance of the gaps explained by the window's transmittance is identical from one day to another. In fact, as figure 12 only shows the relative influence of each parameter, to ensure a better analysis, *var*($\Delta T$) should be taken into account (figure 10). According to this figure, the effect of the transmittance of the window is higher the $8^{th}$ day.

## Results interpretation :

Physically, ($\rho$ , $C_p$, $e$) of a material represent its thermal capacity whereas ($e$, $\lambda$) represent its thermal resistance. So, one can note that it's the thermal capacities of the cement fibre of the roof and weight concrete in the floor that have an influence on indoor air temperature whereas, concerning the polyurethane of the walls and the door, it's their thermal resistance that have an effect (low) on indoor air temperature. This result is not surprising as weight concrete has a high thermal capacity.

The fact that window's properties are the most important factors is not surprising, as it's the first heat source since the cell test 's walls are isolated (polyurethane, polystyrene). The preponderance of outdoor radiative heat transfer coefficient with the fictive sky temperature (*Hrc*) shows that great care should be taken when outdoor radiative heat transfer are linearized.

# Conclusion

In this paper, we introduced a method to perform sensitivity analysis. An application in building thermal simulation allowed to find useful results and showed that among the whole set of parameters, only a few are really influential. Moreover, the fact that some parameters can be interpreted physically reinforces the reliability of the method. SA allows a diagnostic of the building, showing that most important properties belong to the window, the roof and the floor. Thanks to this analysis, we know that in future experimentation (for empirical validation of thermal building simulation code) those parameters should be known accurately or special measurements should be performed to ensure reliable predicted results.

This survey, also shows that SA allows to pinpoint the weaknesses of the model. Indeed, sensitivity of indoor air temperature to the model of radiative heat transfer with the sky incites us to use a higher-level model than the simple linearized one and to measure long-wave heat flux radiation during experiments with a pyrgeometer.

## Acknowledgements

The authors are indebted to Pr. J.P.C Kleijnen of *Tilburg University* and Dr J. Neymark of *Neymark & Associates* for their comments on earlier version of the manuscript. The financial contribution of *Conseil Régional de La Réunion* to this study is gratefully acknowledged.

## APPENDIX

| Wall [Area(m²)] | Layer from interior to exterior | *e* (m) | $\lambda$ (W/m.K) | $C_p$ (J/Kg.K) | $\rho$ (kg/m$^3$) |
|---|---|---|---|---|---|
| East[8], South[7.36], West[8], North[6] | Cement fibre | 0.007 | 0.95 | 1003 | 1600 |
| | Polyurethane | 0.05 | 0.03 | 1380 | 45 |
| | Cement fibre | 0.007 | 0.95 | 1003 | 1600 |
| Roof[9] | Cement fibre | 0.007 | 0.95 | 1003 | 1600 |
| | Polyurethane | 0.05 | 0.03 | 1380 | 45 |
| | Sheet Steel | 0.005 | 163 | 904 | 2787 |
| Floor[9] | Concrete slabs | 0.1 | 0.16 | 653 | 2100 |
| | Polystyrene | 0.5 | 0.04 | 1380 | 25 |
| | Weight concrete | 0.12 | 1.75 | 653 | 2100 |
| Door (North)[2] | Wood | 0.18 | 0.11 | 1500 | 600 |

| | | |
|---|---|---|
| Window (South)[0.64] | K (W/m².K) | 5 |

*Table 2 : Conductive properties of the test cell*

| Walls | $\alpha_i$ | $\alpha_o$ | Hri (W/m².K) | Hre (W/m².K) | Hrc (W/m².K) |
|---|---|---|---|---|---|
| all the walls + doors | 0.6 | 0.3 | 4.5 | 5.7 | 4.7 |

| | | |
|---|---|---|
| Window (South) | $\tau$ | 0.8 |

*Table 3 : Radiative exchange coefficient properties of the test cell*

| Walls | Hci (W/m².K) | Hco (W/m².K) |
|---|---|---|
| all the walls + doors + window | 5 | 11.7 |

*Table 4 : Convective exchange coefficient*